\documentclass[epj,preprintnumbers,amsmath,amssymb,floatfix,10pt,prd,superscriptaddress,nofootinbib]{revtex4}
\usepackage{graphicx}
\usepackage{epsfig}
\usepackage{bm}
\usepackage{amsfonts}
\usepackage{subfigure}
\usepackage{color}
\usepackage{amsmath}
\usepackage{amssymb}
\usepackage{graphicx}
\usepackage{subfigure}
\usepackage{graphicx,epsfig} 
\usepackage{dcolumn}
\usepackage{bm}
\usepackage{caption}

\begin{document}


\title{ Effects   of    low  anisotropy   on interacting holographic  and  new agegraphic  scalar fields models of dark energy}

\author{H. Hossienkhani}
\email{hossienhossienkhani@yahoo.com}
\affiliation{Hamedan Branch, Islamic Azad University, Hamedan, Iran}
\author{A. Aghamohammadi}
\email{a.aqamohamadi@gmail.com,or,a.aghamohamadi@iausdj.ac.ir }
\affiliation{Sanandaj Branch, Islamic Azad University, Sanandaj, Iran}
\author{A. Jafari}
\email{Jafari_manesh@yahoo.com}
\affiliation{Hamedan Branch, Islamic Azad University, Hamedan, Iran}
\author{S.W. Rabieei}
\email{wrabiei@gmail.com, }
\affiliation{Sanandaj Branch, Islamic Azad University, Sanandaj, Iran}
\author{A. Refaei}
\email{abr412@gmail.com,or,A.Refaei@iausdj.ac.ir }
\affiliation{Sanandaj Branch, Islamic Azad University, Sanandaj, Iran}

\date{\today}
\begin{abstract}
A spatially homogeneous and anisotropic Bianchi type I universe is studied with the
interacting holographic and new agegraphic   scalar fields  models of dark energy.
 Given, the framework of the anisotropic model, both the dynamics and potential of these scalar field models according to the evolutionary behavior of  both dark energy models are reconstructed.   We also investigate the  cosmological  evolution of interacting  dark energy models, and compare it with observational data and   schematic diagram.
 In order to do so, we focus on observational determinations of the expansion history $H(z)$. Next,  we evaluate effects of anisotropy on various topics, such as    evolution
of the growth of perturbations in the linear regime,   statefinder diagnostic,    Sandage-Loeb (SL) test and   distance modulus  from    holographic and new agegraphic
dark energy models and compare the results with standard FRW and $\Lambda$CDM and $w$CDM models.
 Our numerical result show the effects of the interaction and anisotropy on the evolutionary behavior the    new agegraphic scalar field models.\\

\end{abstract}

\maketitle

{\bf{keywords}:}{ Anisotropic universe; Statefinder diagnostic; Sandage-Loeb test; Holographic dark energy; New agegraphic;  Scalar field}

\section{Introduction}
\label{sec1}

Most remarkable observational discoveries in type-Ia supernovae (SN Iae) prevail that the universe has confirmed  by two
dark components containing dark energy (DE) and dark matter (DM)  \cite{1, 2}.
DE, an unknown energy with large negative pressure, is \textbf{employed} to  explain the present cosmic accelerating expansion  of the universe. The simplest
candidate for DE is Einstein's  cosmological constant ($\Lambda$) which is equivalent to the vacuum energy density in the universe and produces negative pressure
with equation of state (EoS) parameter  $\omega_\Lambda=-1$ since it fits the observational data. However, this model suffers from two fundamental problems, which are the fine-tuning problem and the cosmic coincidence problem. \\
Recently, considerable interest has been stimulated in explaining the observed DE by the holographic dark energy (HDE) model.   The  HDE is one of interesting DE candidates which was proposed based on the holographic principle    \cite{7,8,9,a9}. According to the holographic principle, the entropy of a system \textbf{mismatch}  with its volume, but with its surface area \cite{9a}.     The HDE was first proposed in  \cite{10} following the line of  \cite{11} where the infrared cut-off is taken to be the size of event horizon for DE.
Ref. \cite{11} purposed that the DE should obey this principle, thus its energy density has an upper limit and the {\bf fine-tuning} problem for the cosmological constant is eliminated.
Its framework is the black hole thermodynamics \cite{12} and the correspondence  of the UV cut-of of a quantum field theory, which gives rise to the vacuum energy,
with the  distance of the theory \cite{14,15}. In the following,  Li \cite{10} and Hsu \cite{15a}  suggested  the model   $\rho_\Lambda=3c^2L^{-2}$, where $c$  is a positive constant and $L$ is the IR cut-off radius.  If in a cosmological context we saturate this inequality by doing  $L=H^{-1}$  \cite{16}, where $H$ is the Hubble parameter, we obtain a model of holographic nature for the  density of DE.   Ref. \cite{17}  was presented a generalized and restored HDE in the braneworld context.
The holographic phantom energy density grows rapidly and dominates the  late time expanding phase, helping realize a cyclic universe scenario in Ref. \cite{17a}.\\
Another plan to explore the nature of DE, dubbed agegraphic DE (ADE), has been proposed  \cite{18} which explain the acceleration of the
universe expansion with the gravitational effect in general relativity.
In the ADE models the age of the universe is taken  the length measure instead of the horizon
distance, so the causality problem that appears in the HDE model can be avoided \cite{18a}. After  introducing  the   ADE model  by Cai \cite{18}, a new model of agegraphic DE (NADE) was proposed in Ref. \cite{19}, while the time scale is chosen to be the conformal time $\eta$ instead of the age of the universe i.e. $T$. The  ADE \textbf{go through a}  the difficulty to characterize the matter dominated era  \cite{19} while the NADE  resolved this subject \cite{20}.    {\bf  An alternative proposal for DE is the dynamical DE scenario which is often realized by scalar field mechanism. It suggests that the energy form with negative pressure is provided by a scalar
field evolving down a proper potential. Some works have been investigated on the reconstruction of the scalar field in NADE models. For instance we refer to  this case as  agegraphic quintessence  \cite{21,22,28}. Also, Ref. \cite{29} focused on the  issue of age problem in the NADE model and determined  the age of the universe in the NADE model by fitting the observational data}.\\
A Bianchi type I (BI) universe, being the straightforward generalization of the flat FRW universe, is of interest because it is one of the simplest models of a non-isotropic
universe exhibiting a homogeneity and spatial flatness. In this case, unlike the FRW universe which has the same scale factor for three spatial directions, a BI universe has
a different scale factor for each direction. This fact introduce a non-isotropy to the system.
 The possible effects of anisotropy in the early universe have been  investigated with  BI  models from different points of view \cite{30,33,34, 35}.
 Therefore, we establish a  correspondence between the interacting holographic and new agegraphic DE  scalar field in an anisotropic universe.\\
The outline of this work is as follows. In the next section, a brief review of the general formulation of the field equations in a BI metric are discussed, then, interacting DE with  cold DM (CDM) in a non-isotropic BI universe is studied. The Sec. \ref{sec3}  is concerned  with interacting  HDE with the quintessence,  tachyon and K-essence scalar field in an anisotropic  universe.   The Sec. \ref{sec4}, is related to  establish the correspondence between the  model of interacting NADE and the quintessence,  tachyon and K-essence DE in   BI model. In   Sec.  \ref{sec5} we study effects of anisotropy on the $H(z)$ data,  {\bf Sandage-Loeb (SL) test},  the  statefinder diagnostic,  the linear evolution of perturbations and distance modulus with both HDE and NADE  models   and compares it with the $\Lambda$CDM, $w$CDM and FRW models.
    We summarize our results in last section.
\section{General framework}
\label{sec2}
Bianchi cosmologies are spatially homogeneous but not necessarily isotropic. Here we will consider BI cosmology. The metric of this model is given by
\begin{equation}\label{1}
ds^2=dt^{2}-A^{2}(t)dx^{2}-B^{2}(t)dy^{2}-C^{2}(t)dz^{2},
\end{equation}
where the metric functions, $A, B, C$, are merely  functions of time, $t$.   However, the underlying Lie algebra of the isometry group of the BI metrics is completely different \cite{35a}.   The contribution
of the interaction with the matter fields is given by the energy momentum tensor, which is defined as
\begin{eqnarray}\label{4}
T^{\mu}_{\nu}=diag[\rho,-\omega\rho,-\omega\rho,-\omega\rho],
\end{eqnarray}
where $\rho$ and $\omega$ represent the energy density and EoS parameter, respectively.  The field equations for the axially symmetric BI metric are \cite{36,37,38}:
\begin{eqnarray}
3H^{2}-\sigma^{2}&=&\frac{1}{M_p^2}(\rho_{m}+\rho_{\Lambda}),   \label{12} \\
3H^2+2\dot{H}+\sigma^{2}&=&-\frac{1}{M_p^2}\left(p_{m}+p_{\Lambda}\right), \label{13}
\end{eqnarray}
where  $M_p^2=1/(8\pi G)$, $\rho_{\Lambda}$ and $p_{\Lambda}$ are the Planck mass, the energy density and pressure of  DE, respectively, and  $a=(ABC)^{\frac{1}{3}}$ is the scale factor, and
$\sigma^2=1/2\sigma_{ij}\sigma^{ij} $  in which
$\sigma_{ij}=u_{i,j}+\frac{1}{2}(u_{i;k}u^{k}u_{j}+u_{j;k}u^{k}u_{i})+\frac{1}{3}\theta(g_{ij}+u_{i}u_{j})$
is the shear tensor $(\sigma_{ij}u^j=0, \sigma^i_{~i}=0)$, which describes the rate of distortion of the
matter flow, and $\theta=3H=u^{j}_{;j}$ is the  scalar expansion, where $u^j$ is 4-velocity.  In a comoving coordinate system, i.e. $(u^i=\delta^i_0)$.
From the metric  Eq. (\ref{1}),  and considering the comoving frame,  {\bf hence,} the components of the average Hubble parameter and the shear tensor
are given by \cite{36}
\begin{eqnarray}
H&=&\frac{1}{3}(\frac{\dot{A}}{B}+\frac{\dot{B}}{B}+\frac{\dot{C}}{C}),   \label{14} \\
\sigma^{2}&=&3H^2-(\frac{\dot{A}\dot{B}}{AB}+\frac{\dot{B}\dot{C}}{BC}+\frac{\dot{A}\dot{C}}{AC}). \label{15}
\end{eqnarray}
By using  Eq. (\ref{12}),   the dimensionless density parameter of BI can also be defined as usual
\begin{eqnarray}\label{16}
\Omega_{m} &=&\frac{\rho_{m}}{\rho_{cr}}, \quad  \Omega_{\Lambda} =\frac{\rho_{\Lambda}}{\rho_{cr}},
\end{eqnarray}
where the critical energy density is $\rho_{cr}=3M_p^2H^2$. Thus, the first BI equation can be rewritten as
\begin{equation}\label{17}
\Omega_m+\Omega_{\Lambda}=1-\frac{\sigma^2}{3H^2}.
\end{equation}
The above equation  shows that the sum of the energy density parameters approaches 1 at late times \textbf{if the shear tensor tend zero} .  Hence, at the late times the universe becomes flat,\textbf{i.e.}  for sufficiently large time, this model predicts that the anisotropy of the universe will damp out and universe will become
isotropic. \textbf{By the way ,} in the early universe i.e. during the radiation and matter
dominated era the universe \textbf{have been} anisotropic and the universe approaches to isotropy \textbf{upon}  DE starts to dominate the energy density of the universe.  When the isotropic is assumed, $\Omega_m+\Omega_{\Lambda}=1$, and the model has only one free parameter, $\Omega_{\Lambda}$.
We shall take that the shear scalar can be described based on the average Hubble parameter, $\sigma^2=\sigma_0^2H^2$, where $\sigma_0^2$ is a constant. So, Eq. (\ref{17}) lead to
\begin{eqnarray}\label{17a}
\Omega_m+\Omega_{\Lambda}=1-\Omega_{\sigma0},\quad ~with ~~\Omega_{\sigma0}=\frac{\sigma_0^2}{3},
\end{eqnarray}
where $\Omega_{\sigma0}$ is the anisotropy parameter.
   We assume that the DE is coupled with DM, in such  way that
total energy-momentum is still conserved. In the flat BI cosmology with a scale factor $a$, the continuity equations for both components are   \cite{38a}
\begin{eqnarray}\label{18}
\dot{\rho}_\Lambda+3H\rho_\Lambda(1+\omega_\Lambda)=-Q,\\
\dot{\rho}_m+3H\rho_m=Q,\label{19}
\end{eqnarray}
where $\omega_\Lambda=p_\Lambda/\rho_\Lambda$ is the EoS parameter of the interacting DE
and $Q$ is the interaction term. The case $Q>0(Q<0)$ corresponds to  either DE transformation into DM and vice versa of the model.  Following  \cite{14,15},   we shall assume $Q=3b^2H(\rho_m+\rho_\Lambda)$ with the coupling constant  $b^2$.  In the following, we will find a form for the function  $V(\phi)$, which is able to reconstruct HDE and NADE models  as following.
\section{HDE with Hubble radius as IR cutoff  in BI  model}
\label{sec3}
First we will consider the size $L$ as the Hubble radius $H^{-1}$, thus the energy density
for the DE becomes  \cite{7,41,42}
\begin{eqnarray}\label{21}
\rho_\Lambda=3c^2 M_p^2 H^2,
\end{eqnarray}
where $c^2$ is a constant. As it is well known, different possibilities have been tried to
varying degrees of success, namely, the particle horizon \cite{43}, the future event horizon
\cite{10,14} and the Hubble horizon \cite{16}. Hence,  we set the Hubble radius as the infrared cutoff $L=H^{-1}$ \cite{44}. \textbf{The} problem of taking Hubble radius,\textbf{is that} the DE results as pressureless, since $\rho_\Lambda$ scales like matter energy density $\rho_m$ with the scale factor $a$ as $a^{-3}$, therefore producing a decelerated expansion.\textbf{But}
at this point our system of equations is not closed and we still have freedom to choose. In order to accelerate the universe more, we have added the interaction as well as anisotropy, as it can be seen in equation  (\ref{25}). This is a new
model to explore the physical mechanism responsible for the current acceleration of the universe.
 Therefor in the present work, we have tried to compensate the effects, (due to the Hubble's first radius that lead to a deceleration of the universe) using interaction and anisotropy. For $b^2=\Omega_{\sigma0}=0$   this result is consistent with the Refs. \cite{15a,42,44,45,451}. Now by taking the time derivative of relation (\ref{21}) and using the BI equation we find
\begin{eqnarray}\label{24}
\dot{\rho}_{\Lambda}=2\rho_{\Lambda}\frac{\dot{H}}{H}=-3\frac{Hc^2 \rho_{\Lambda}}{1-\Omega_{\sigma0}}(1+\omega_{\Lambda}+r),
\end{eqnarray}
where $r=\rho_m/\rho_\Lambda=(1-\Omega_{\sigma0}-\Omega_{\Lambda})/\Omega_{\Lambda}$  is the energy density ratio.   Substituting Eq. (\ref{24}) into   (\ref{18}), the EoS parameter for interaction HDE   is given by
\begin{equation}\label{25}
\omega_{\Lambda}=-\frac{b^2(1-\Omega_{\sigma0})^2}{c^2(1-c^2-\Omega_{\sigma0})}.
\end{equation}
This relation shows that the EoS parameter {\bf depends on the $b^2$,  $c$ and $\Omega_{\sigma0}$ parameters}. Therefore,    $\omega_\Lambda$  becomes   a constant.  In the
absence of interaction between HDE and CDM, $b^2=0$ we have
$\omega_{\Lambda}=0$, which is that matter dominant. It may be pointed out here that if one sets $\Omega_{\sigma0}=0$, Eq. (\ref{25})  reduce to that
obtained for a flat FRW universe \cite{42}.\\
Now, we intend to categorize our study into three types of the model, i.e. quintessence, tachyon and K-essence models in non-isotropic universe.
\subsection{Quintessence reconstruction of HDE in BI}
\label{subsec1}
We consider a scalar field (the quintessence field)  $\phi$, evolving in a potential $V(\phi)$.  The density and pressure of quintessence are given by
\begin{eqnarray}\label{27}
\rho_\phi&=&\frac{1}{2}\dot{\phi}^2+V(\phi),\\
p_\phi&=&\frac{1}{2}\dot{\phi}^2-V(\phi).\label{28}
\end{eqnarray}
Quintessence is usually parameterized by its equation of
state $\omega_\phi=p_\phi/\rho_\phi$. Looking at Eqs. (\ref{27}) and  (\ref{28}), we
see immediately that $\omega_\phi\geq-1$. The scalar field potential and the corresponding kinetic energy of the field are obtained from Eqs. (\ref{27}) and  (\ref{28}),
which are \cite{21}
\begin{eqnarray}\label{29}
V(\phi)&=&\frac{1-\omega_\phi}{2}\rho_\phi,\\
\dot{\phi}^2 &=&(1+\omega_\phi)\rho_\phi,\label{30}
\end{eqnarray}
where $\omega_\phi=p_\phi/\rho_\phi$.  For the accelerated expansion of the universe, the EoS parameter for quintessence must be less than $-1/3$.
By means of $\rho_\phi=\rho_\Lambda$ and $\omega_\phi=\omega_\Lambda$ and   substituting  Eqs. (\ref{21}) and  (\ref{25})  into  (\ref{30})  one can get
\begin{eqnarray}\label{31}
\phi=c M_p  \sqrt{3(1-\frac{b^2(1-\Omega_{\sigma0})^2}{c^2(1-c^2-\Omega_{\sigma0})})} \ln a.
\end{eqnarray}
 using Eqs. (\ref{12}),    (\ref{16}), (\ref{21})   and (\ref{25}) we obtain
\begin{equation}\label{32}
\frac{\dot{H}}{H^2}= -\frac{3}{2}\left(1-\frac{b^2(1-\Omega_{\sigma0})}{1-c^2-\Omega_{\sigma0}}\right).
\end{equation}
Integration of  Eq. (\ref{32}), the scaler  field  (\ref{31}) can be written
\begin{eqnarray}\label{35}
\phi=\frac{2c M_p}{\sqrt{3} } \frac{  \sqrt{1-\frac{b^2(1-\Omega_{\sigma0})^2}{c^2(1-c^2-\Omega_{\sigma0})}}}{1-\frac{b^2(1-\Omega_{\sigma0})}{1-c^2-\Omega_{\sigma0}} }\ln t,
\end{eqnarray}
and the potential
\begin{eqnarray}\label{36}
V(\phi)=\frac{2c^2M_p^2}{3}\bigg( \frac{1+\frac{b^2(1-\Omega_{\sigma0})}{1-c^2-\Omega_{\sigma0}}}{\left(1-\frac{b^2(1-\Omega_{\sigma0})}{1-c^2-\Omega_{\sigma0}} \right)^{2}}\bigg) exp\bigg(-\frac{\sqrt{3}\phi}{cM_p}
\frac{1-\frac{b^2(1-\Omega_{\sigma0})}{1-c^2-\Omega_{\sigma0}}}{\sqrt{1-\frac{b^2(1-\Omega_{\sigma0})^2}{c^2(1-c^2-\Omega_{\sigma0})}}}\bigg).
\end{eqnarray}
From the solution of    (\ref{32}), it is readily {\bf seen that} the scale factor is $a=t^{\frac{2}{3}\left(1-\frac{b^2(1-\Omega_{\sigma0})}{1-c^2-\Omega_{\sigma0}} \right)^{-1}}$.  {\bf It is clear}  that   an accelerated universe, with $\ddot{a}>0$ is yielded   if  $b^2>(1-c^2-\Omega_{\sigma0})/3(1-\Omega_{\sigma0})$.
\subsection{Tachyon reconstruction of HDE in BI}
\label{subsec2}
On the other hand, among the various candidates to explain the accelerated expansion, the
rolling tachyon condensates in a class of string theories may have interesting cosmological consequences \cite{46a,46b}. Also, the tachyon field is originated from the D-brane action in string theory \cite{46c,46d,46e,46h}. It can be introduced by a simple manner as follows. In a flat BI background the energy density  and the
pressure density  are given by \cite{46,47,48}
\begin{eqnarray}\label{37}
\rho_T&=&-T_0^0=\frac{V(\phi)}{\sqrt{1-\dot{\phi}^2}},\\
p_T&=&T_i^i=-V(\phi)\sqrt{1-\dot{\phi}^2}.\label{38}
\end{eqnarray}
 The EoS of the tachyon is given by
\begin{eqnarray}\label{39}
\omega_T=\dot{\phi}^2-1.
\end{eqnarray}
 In the following, one can express   $\phi$ in terms of  $b^2$, $c^2$ and $\Omega_{\sigma0}$.  Hence,   equating (\ref{25}) with (\ref{39}), i.e. $\omega_\Lambda=\omega_T$,  gives
\begin{equation}\label{41}
\phi(t)=\sqrt{1-\frac{b^2(1-\Omega_{\sigma0})^2}{c^2(1-c^2-\Omega_{\sigma0})}} t.
\end{equation}
Integrating of    Eq. (\ref{32})   and  using (\ref{41})  we obtain tachyon {\bf the} potential in term of the scalar field
\begin{equation}\label{43}
V(\phi)=\frac{4M_p^2 cb(1-\Omega_{\sigma0})}{3\phi^{2}\sqrt{1-c^2-\Omega_{\sigma0}}}\frac{1-\frac{b^2(1-\Omega_{\sigma0})^2}{c^2(1-c^2-\Omega_{\sigma0})}}{(1-\frac{b^2(1-\Omega_{\sigma0})}{1-c^2-\Omega_{\sigma0}})^2}.
\end{equation}
In this case the evolution of the tachyon is as $\phi \propto t$. Tachyon potentials, which  are not steep compared to $V(\phi)\propto\phi^{-2}$ lead to an accelerated expansion  \cite{49}.
\subsection{K-essence reconstruction of HDE in BI}
\label{subsec3}
The scalar field model named K-essence is also employed to describe the observed acceleration of the cosmic expansion. This kind of models is characterized by non-canonical kinetic energy terms, and are described
by a general scalar field action, which is a function of $\phi$ and $\chi=\dot{\phi}^2/2$, and is given by \cite{49a,50}
\begin{equation}\label{44}
S=\int d^4x\sqrt{-g}P(\phi,\chi),
\end{equation}
where  $P(\phi,\chi)$  is an arbitrary function of $\phi$ and its kinetic energy $\chi$. Based on the analysis of the Lagrangian density, $‎\mathcal{L}$ ,\textbf{in which,$\mathcal{L}=P,$ } it can be transformed into
\begin{equation}\label{45}
P(\phi,\chi)=f(\phi)(-\chi+\chi^2),
\end{equation}
and the energy density in these models is given by
\begin{equation}\label{46}
\rho(\phi,\chi)=f(\phi)(-\chi+3\chi^2).
\end{equation}
And the  EoS parameter $\omega_K=P(\phi,\chi)/\rho(\phi,\chi)$  is
\begin{equation}\label{47}
\omega_K=\frac{\chi-1}{3\chi-1}.
\end{equation}
This shows that $\omega_K$  does not change  for fixed  $\chi$. Equating  $\omega_K=\omega_\Lambda$ from  Eq. (\ref{25}) and   (\ref{47}),  we get
\begin{equation}\label{48}
\chi=\frac{b^2(1-\Omega_{\sigma0})^2+c^2(1-c^2-\Omega_{\sigma0})}{3b^2(1-\Omega_{\sigma0})^2+c^2(1-c^2-\Omega_{\sigma 0})}.
\end{equation}
The EoS parameter in Eq. (\ref{47}) diverges for $\chi=1/3$.  For the  case of $\chi>1/3$, the condition $\omega_K<-1/3$ leads to a high bound on $\chi<2/3$. The  requirement of   an accelerated expansion gives $1/3<\chi<2/3$ and the cosmological constant limit corresponds to $\chi=1/2$. Using Eq. (\ref{48}) and $\chi=\dot{\phi}^2/2$, one can write the equation for the K-essence field as follows
\begin{equation}\label{49}
\phi(t)=\sqrt{2\frac{b^2(1-\Omega_{\sigma0})^2+c^2(1-c^2-\Omega_{\sigma0})}{3b^2(1-\Omega_{\sigma0})^2+c^2(1-c^2-\Omega_{\sigma0})}}t,
\end{equation}
Note that the field increases with the increment of $t$. From Eqs.   (\ref{46}) and   (\ref{48}) it can be obtained an expression for the k-essence   potential $f(\phi)$ in terms of the $\phi$
\begin{equation}\label{51}
f(\phi)=\frac{4M_p^2}{3\phi^2}\frac{3b^2(1-\Omega_{\sigma0})^2+c^2(1-c^2-\Omega_{\sigma0})}{1-c^2-\Omega_{\sigma0}-b^2(1-\Omega_{\sigma0})},
\end{equation}
where  Eq. (\ref{47}) was used. From Eq. (\ref{49}) we see that  the kinetic energy of K-essence  is not constant and is variable with  cosmic time.
\section{NADE with scalar field in BI  model}
\label{sec4}
 The energy density of the NADE can be written \cite{19,20}
\begin{equation}\label{52}
 \rho_{\Lambda}=\frac{3n^2M_p^2}{\eta^2},
\end{equation}
 where the numerical factor $3n^2$ has been introduced to parameterize some uncertainties, such as the species of quantum fields in the universe, or the effect of curved space-time. The conformal time is given by
\begin {equation}\label{53}
 \eta=\int\frac{dt}{a(t)}=\int \frac{da}{Ha^2}.
\end{equation}
If $\eta$ to be a \textbf{indefinite} integral, \textbf{the integrand give a function plus integral constant} . Thus, we have $\dot{\eta}=1/a$.    The  energy density parameter  of the NADE is now given by
\begin{equation}\label{54}
\Omega_{\Lambda}=\frac{n^2}{H^2\eta^2}.
\end{equation}
Taking the derivative of Eq.  (\ref{52}) with respect to the cosmic time and using  (\ref{54}) we  get
\begin{equation}\label{55}
\dot{\rho}_{\Lambda}=-2H \frac{
\sqrt{\Omega_{\Lambda}}}{na}\rho_{\Lambda}.
\end{equation}
 Inserting this relation into (\ref{18}) we obtain the EoS parameter of the interacting NADE
\begin{equation}\label{56}
\omega_{\Lambda}=-1+\frac{2}{3na}\sqrt{\Omega_{\Lambda}}-\frac{b^2}{\Omega_\Lambda}(1-\Omega_{\sigma0}).
\end{equation}
It is important to note that when $b^2=0$, the interacting DE becomes trivial and Eq. (\ref{56})
reduces to its respective expression in new ADE in general relativity \cite{36}.
 \begin{figure}[tb]
 \includegraphics[width=.35\textwidth]{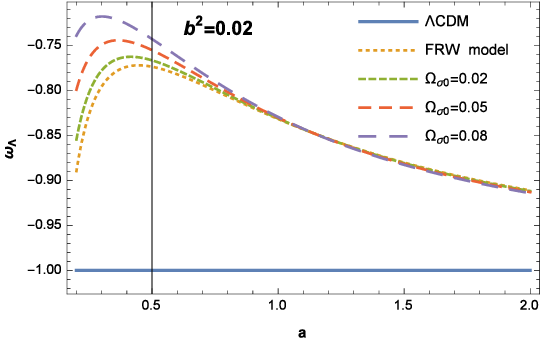}\hspace{1cm}
 \includegraphics[width=.35\textwidth]{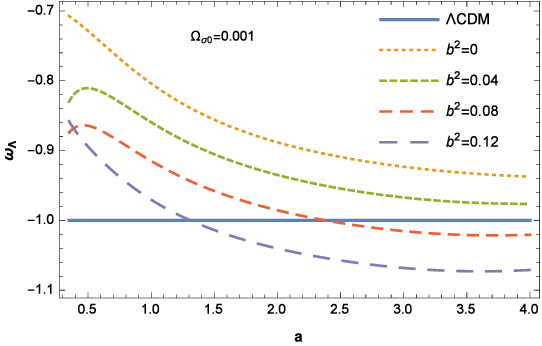}
\caption{The plot shows the evolution of the EoS parameter of NADE, Eq. (\ref{56}), for different the anisotropy energy density parameter  $\Omega_{\sigma0}$ and  the coupling constant $b^2$.   For both cases, we take   $n=2.886$ and $\Omega_{\Lambda}^0=0.72$.}
 \label{fig:1}
 \end{figure}
In this case ($b^2=0$), the present accelerated expansion of our universe can be derived only if
$n>1$ \cite{19}, in addition, from this figure we see that $\omega_{\Lambda}$ of the NADE model cannot cross the phantom divide and the universe has a quintessence region at late time; eventually, for the case of matter dominated era $\omega_\Lambda=-2/3$ whilst $\Omega_\Lambda=n^2a^2/4$ and in the radiation dominated era $\omega_\Lambda=-1/3$ whilst $\Omega_\Lambda=n^2a^2$.
We test this scenario for the interaction between NADE and DM by using some observational results.
 For the comparison with the phenomenological interacting model, in our scenario the coupling between NADE and DM can be expressed  by $b^2$ parameter as in the phenomenological interaction form. In fact, $b^2$ is within the region of the golden supernova data fitting result $b^2=0.00^{+0.11}_{-0.00}$ \cite{14}  and the observed CMB low $l$ data constraint \cite{51}.  In figure 1 we plot the evolution behaviors of EoS  $\omega_{\Lambda}$  of DE, with respect to different  choices of anisotropy energy density parameter $\Omega_{\sigma0}$ and interaction parameter $b^2$ and the $\Lambda$CDM model.
  It clear that the EoS parameter decrease with the scale factor increase  and the effect of various $\Omega_{\sigma0}$ are negligible but as $a\rightarrow 0$, it is clear that increasing of the $\Omega_{\sigma0}$ cause to increase the EoS parameter. Notice that in the   case of $b^2\geq 0.08$, $\omega_\Lambda$ crosses the phantom divide line $\omega_\Lambda=-1$.
 Another best fit data with the NADE model is  $n = 2.886^{+0.084}_{-0.082}$ at the $1\sigma$ level and $n = 2.886^{+0.169}_{-0.163}$ at the $2\sigma$ level \cite{29},   $n = 2.807^{+0.087}_{-0.086}(1\sigma) ^{+0.176}_{-0.170}(2\sigma)$ \cite{52},
for using the SNIa data, it is  $n=2.954^{+0.264}_{-0.245}$ at  $1\sigma$ Confidence Level and   $n=2.716^{+0.111}_{-0.109}$  with SNIa, CMB and Large Scale Structure data \cite{20} and for the non-flat universe $n=2.673^{+0.053+0.127+0.199}_{-0.077-0.151-0.222}$ \cite{52a}.   For the $\Lambda CDM$ cosmology, $\Omega_{m0}=0.274$ is  given by the WMAP five-year observations \cite{30}.
In what follows given to the Eq. (\ref{56}) and its evaluations in the Fig. (\ref{1}), by pick $b^2\neq 0$, taking account  $\Omega_{m0}=0.274$ \cite{30}, $\Omega_{\Lambda 0}=1-(\Omega_{m0}+\Omega_{\sigma0})$,  $\Omega_{\sigma0}=0.001$, $n=2.7$ \cite{20} and  $a=1$ for the present time, Eq. (\ref{56}) gives
\begin{equation}\label{58}
\omega_{\Lambda}=-0.7905-1.3875 b^2,
\end{equation}
where is clear that the phantom EoS $\omega_{\Lambda}<-1$ can be achieved by set $ b^2>0.15$  for the coupling between NADE and CDM.
In the future, where  $a\rightarrow \infty$, $\omega_\Lambda <-1$ for $b^2>0$, i.e.  it may be the $\omega_\Lambda$  crosses the phantom divide line in the presence interacting DM and DE.  We can also obtain the evolution behavior of the DE. Taking the time derivative of $\Omega_{\Lambda}$  in Eq. (\ref{54}) and relation
$\dot{\Omega}_{\Lambda}=H\Omega_{\Lambda}'$, give
 \begin{figure}
 \centering
{\includegraphics[scale=0.55]{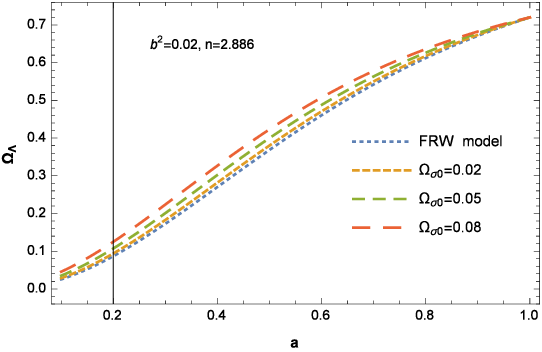}}~~~~
{\includegraphics[scale=0.55]{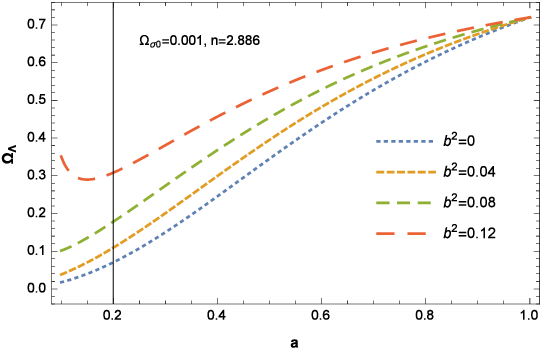}}~~~~
 {\includegraphics[scale=0.55]{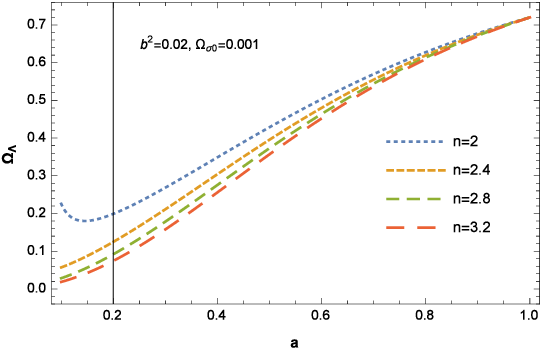}}~~~~
\caption{  (a) $\Omega_\Lambda$   versus    $a$ for  different  $\Omega_{\sigma0}$ with $b^2=0.02$ and $n=2.886$; (b)  $\Omega_\Lambda$    versus    $a$ for  different  $b^2$  with $\Omega_{\sigma0}=0.001$ and $n=2.886$; (c) $\Omega_\Lambda$  versus    $a$ for  different  $n$ with $\Omega_{\sigma0}=0.001$ and $b^2=0.02$. The initial data is same as in Fig. (1). }
\label{fig:2}
\end{figure}
\begin{equation}\label{59}
\Omega_{\Lambda}'=-2\Omega_{\Lambda}\left(\frac{\dot{H}}{H^2}+\frac{\sqrt{\Omega_\Lambda}}{na}\right).
\end{equation}
Taking the derivative of  the BI equation (\ref{12}) with respect to the cosmic time, and using
Eqs. (\ref{17}), (\ref{18}), (\ref{52}) and (\ref{54}), it is easy to find
\begin{equation}\label{60}
\frac{\dot{H}}{H^2}=-\frac{3}{2}\bigg(1-b^2-\frac{\Omega_\Lambda}{1-\Omega_{\sigma0}}+\frac{2}{3na}\frac{\Omega_\Lambda^{\frac{3}{2}}}{1-\Omega_{\sigma0}}\bigg).
\end{equation}
Combining Eqs. (\ref{59}) and (\ref{60}),  we have the equation of motion, a differential equation, for $\Omega_{\Lambda}$
\begin{equation}\label{61}
\Omega_{\Lambda}'=3\Omega_{\Lambda}\left(1-b^2-\frac{\Omega_\Lambda}{1-\Omega_{\sigma0}}-(1-\frac{\Omega_\Lambda}{1-\Omega_{\sigma0}})\frac{2\sqrt{\Omega_\Lambda}}{3na}\right) .
\end{equation}
As shown in Ref. \cite{19}, the parameters  $n$ and $\Omega_{mo}$  are not independent of each other. If one takes  them as free
parameters, the NADE model will become  confused. Therefore the initial conditions in NADE should be taken at the
early times.  Alike Ref. \cite{19} proposed the initial condition for the case of ADE,  $\Omega_{\Lambda}=n^2a^2/4$, at $z=1/a-1=2000$, so Eq. (\ref{61}) can be
numerically solved. In Fig. (2) we have depicted the evolution of $\Omega_{\Lambda}$ as a function of scale
factor for various values of the free parameters for this model. {\bf Figure (2a) show the effects of the anisotropic on the evolutionary behavior the NADE model, it clear that $\Omega_{\Lambda}$ increase with increase of anisotropic parameter density at the smaller scale factor, but at the $a=1$, the effect of anisotropic parameter density is negligible,that is in agreement with observational data, i.e. the present universe is close to  homogeneous and isotropic flat universe.  Figures (2b) and (2c)  evolution of $\Omega_{\Lambda}$ are plotted for different choices of $b^2$ and $n$,it is clear,  the $b^2$  parameters increase and the $n$ decrease  cause the $\Omega_{\Lambda}$ increase in the smaller scale factor but in the $a=1$, the curves are coincide, that is also in agreement with the observational data .
 On the other word, the evolution behaviour of $\Omega_{\Lambda}$ is as following:
\begin{description}
\item[(I)]
For the cases of $b^2\neq 0.12$ and $n\neq 2$, we can see that at the early time $\Omega_{\Lambda}\rightarrow0$, and
hence the universe always behaves as an Einstein-de Sitter (EdS) cosmology, while at
the  late time $\Omega_{\Lambda}\rightarrow 1$,  that is the NADE dominates as expected.
\item[(II) ]
Finally, the plot illustrates the $\Omega_{\Lambda}$  increase with the $a$ increase.
 \end{description}}

\subsection{Quintessence reconstruction of NADE in BI}
\label{subsec4}
Now we suggest a correspondence between the NADE and quintessence scalar field namely,
we identify $\rho_\phi$ to be $\rho_\Lambda$. Using relation $\rho_\phi=\rho_\Lambda=3M_p^2H^2\Omega_\Lambda$ and substituting  Eqs. (\ref{54}) and    (\ref{56}) into  (\ref{29}) and (\ref{30}) one can readily  find the kinetic energy term and the  potential term as
 \begin{eqnarray}
\dot{\phi^2}&=&M_p^2H^2\left(\frac{2}{na}\Omega^{\frac{3}{2}}_{\Lambda}-3b^2(1-\Omega_{\sigma0})\right), \label{63}\\
V(\phi)&=&M_p^2H^2\left(3\Omega_{\Lambda}+\frac{3b^2}{2}(1-\Omega_{\sigma0})-\frac{\Omega^{\frac{3}{2}}_{\Lambda}}{na}\right).\label{64}
\end{eqnarray}
\begin{figure}[h]
\includegraphics[width=0.29\textwidth]{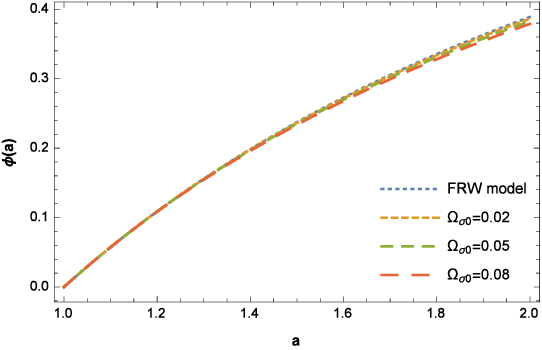}\hspace{.5cm}
 \includegraphics[width=0.3\textwidth]{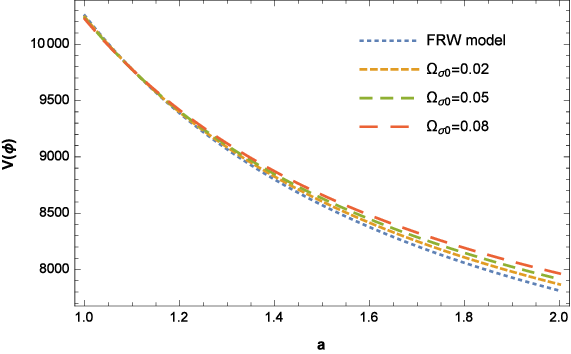}\hspace{.5cm}
 \includegraphics[width=0.3\textwidth]{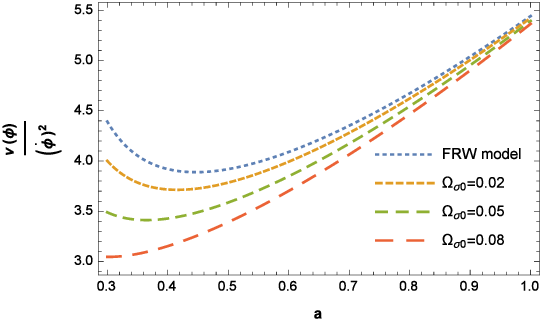}
\caption{{\small\label{fig:3}Left panel: The plot shows the evolution of the NADE quintessence scalar {\bf field}, Eq. (\ref{65}), for four different value of  the anisotropy energy density parameter   $\Omega_{\sigma0}$. Middle panel: The plot shows the NADE quintessence potential, Eq. (\ref{64}), versus the scale factor for different  $\Omega_{\sigma0}$. Right panel:  The plot shows the NADE quintessence $\frac{V(\phi)}{\dot{\phi}^2}$,  for   different value of $\Omega_{\sigma0}$. Auxiliary parameters are $\Omega_\Lambda^0=0.72$, $b^2=0.02$,  $n=2.886$ and $\phi(1)$=0.}}
\end{figure}
From definition  $\dot{\phi}=H\phi'$, one can rewrite Eq. (\ref{63})  as
\begin{equation}\label{65}
\phi'=M_p\left(\frac{2}{na}\Omega^{\frac{3}{2}}_{\Lambda}-3b^2(1-\Omega_{\sigma0})\right)^{\frac{1}{2}}.
\end{equation}
Therefore, we have established an interacting new agegraphic quintessence DE model and reconstructed the potential of the agegraphic quintessence as well as the dynamics of scalar field in an anisotropic universe. Basically, from Eqs.  (\ref{61}) and   (\ref{65}) one can derive
$\phi=\phi(a)$ and then combining the result with   (\ref{64}) 
\textbf{ahieve} $V=V(\phi)$. Unfortunately, the analytical form of the potential in terms of the new agegraphic quintessence field cannot be determined due to the complexity of the equations involved. However, we
can obtain it numerically.  The evolution of the NADE quintessence scalar {\bf field} and the potential $V(\phi)$ for four different values of $\Omega_{\sigma0}$ is plotted in  left and middle  panel of Fig. 3. We see that the differences in the \textbf{diagrams}   of $\phi$ are very little. \textbf{It is clear, } $\phi(a)$
increases and $V(\phi)$ decreases as universe expands and the curves are shifted to the smaller (bigger) values of  $ \phi (a) (V(\phi))$ with increasing the $\Omega_{\sigma0}$. Also, we see from these figures that the cosmic evolution trends are quite similar for these four   anisotropy energy density parameters.
The evolution of $V(\phi)/\dot{\phi}^2$ versus $a$ is illustrated in right Fig. (3). It illustrate fraction of potential per kinetic energy increase with the $a$ increase and its magnitude increase with the $\Omega_{\sigma0}$ decrease for the smaller $a$,  that is consist of with slow-roll approximation on inflation era for the smaller $a$ at early universe.
\subsection{Tachyon reconstruction of NADE in BI}
\label{subsec5}
Next, we reconstruct the new agegraphic tachyon DE model, connecting the tachyon scalar field with the NADE in BI universe. Using Eqs. (\ref{54}) and (\ref{56}) one can easily show that the tachyon potential and kinetic energy term take the following form
 \begin{eqnarray}
V(\phi)&=&3M_p^2  H^2\Omega_\Lambda \left(1-\frac{2}{3na}\sqrt{\Omega_\Lambda}+\frac{b^2}{\Omega_\Lambda}(1-\Omega_{\sigma0}) \right)^{\frac{1}{2}},\label{67}\\
\dot{\phi}&=&\left(\frac{2}{3na}\sqrt{\Omega_\Lambda}-\frac{b^2}{\Omega_\Lambda}(1-\Omega_{\sigma0})\right)^{\frac{1}{2}}. \label{68}
\end{eqnarray}
Therefore the evolution behavior of the tachyon field can be obtained by integrating the above
equation
\begin{equation}\label{70}
\phi(a)-\phi(1)= \int^{a}_{1} \frac{1}{Ha}\sqrt{\frac{2}{3na}\sqrt{\Omega_\Lambda}-\frac{b^2}{\Omega_\Lambda}(1-\Omega_{\sigma0})}da,
\end{equation}
\begin{figure}[h]
 \includegraphics[width=0.35\textwidth]{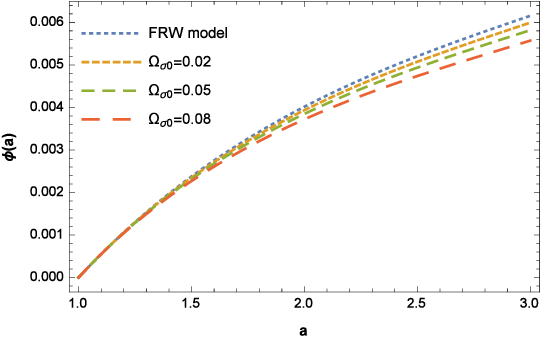}\hspace{1cm}
 \includegraphics[width=0.35\textwidth]{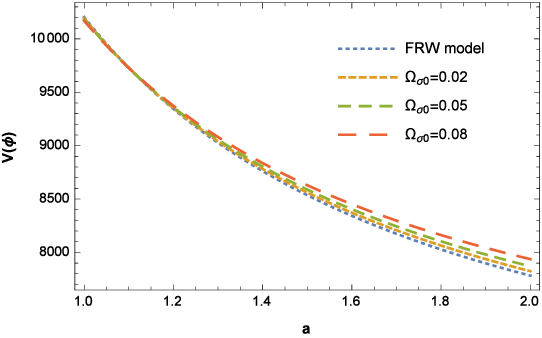}
 \caption{\label{fig:4}Right panel: The evolution of the NADE tachyon scalar {\bf field}, Eq. (\ref{70}), for different the anisotropy energy
density parameter   $\Omega_{\sigma0}$. Right panel: The NADE tachyon potential, Eq. (\ref{67}), versus the scale factor for different  $\Omega_{\sigma0}$. Auxiliary parameters as in Fig.  (3).}
\end{figure}
where $\Omega_\Lambda$ is given by Eq.  (\ref{61}). The evolutionary form of the tachyon field and the reconstructed tachyon potential $V(\phi)$ are plotted in Fig.
(4),\textbf{ for different value of the anisotropy  density parameter, in which   again  have taken $\phi(a_0=1)=0$}  for the present time. $\Omega_{\sigma0}$. From this figure we find out that $\phi$ increases with time
while the potential $V(\phi)$ becomes steeper with decreasing $\Omega_{\sigma0}$. This behavior is in agreement with the scaling solution $V(\phi)\propto\phi^{-2}$ obtained for the tachyon {\bf field} corresponding to the power law expansion \cite{49}. Again from right panel of  Fig. (4) we find out the reconstructed scalar field has the  same  potential  as  the quintessence case.

\subsection{K-essence reconstruction of NADE in BI}
\label{subsec6}
 \begin{figure}
 \includegraphics[width=.35\textwidth]{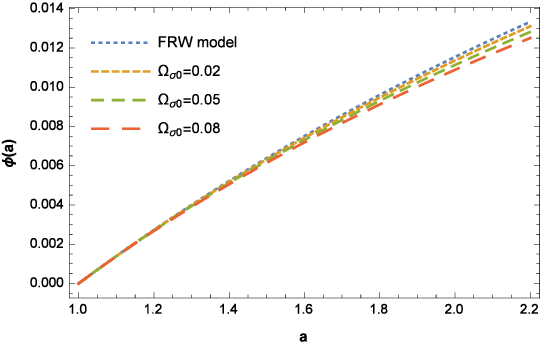}\hspace{1cm}
 \includegraphics[width=0.35\textwidth]{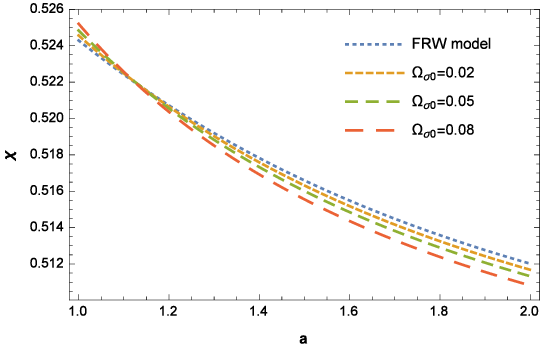}
\caption{Right panel: The K-essence scalar field as function of the scale factor for the variety  values of  the $\Omega_{\sigma0}$. Left panel: The evolution of the new agegraphic K-essence kinetic energy  $\chi=\dot{\phi}^2/2$    for different anisotropy constant  $\Omega_{\sigma0}$. Auxiliary parameters as in Fig. (3).}
 \label{fig:5}
 \end{figure}
The K-essence scalar field model is also an interesting attempt to explain the origin of DE using string theory. Equating $\omega_K$ with the EoS parameter of NADE (\ref{56}) one finds
\begin{equation}\label{71}
\chi=\frac{2-\frac{2}{3na}\sqrt{\Omega_{\Lambda}}+\frac{b^2}{\Omega_\Lambda}(1-\Omega_{\sigma0})}
{4-\frac{2}{na}\sqrt{\Omega_{\Lambda}}+\frac{3b^2}{\Omega_\Lambda}(1-\Omega_{\sigma0})}.
\end{equation}
Using Eq. (\ref{71}) and $\dot{\phi}^2=2\chi$, we obtain the new agegraphic K-essence scalar field as
\begin{equation}\label{72}
\phi(a)-\phi(1)= \int^{a}_{1} \frac{1}{Ha}\sqrt{\left(\frac{4-\frac{4}{3na}\sqrt{\Omega_{\Lambda}}+\frac{2b^2}{\Omega_\Lambda}(1-\Omega_{\sigma0})}
{4-\frac{2}{na}\sqrt{\Omega_{\Lambda}}+\frac{3b^2}{\Omega_\Lambda}(1-\Omega_{\sigma0})} \right) }da.
\end{equation}
where we take $a_0=1$ for the present time.  The evolution of the field and the reconstructed K-essence kinetic energy $\chi(\phi)$ are plotted in Fig. (5), where we have taken $\phi (a_0=1)= 0$ for simplicity.  Note that in left Fig. (5)  the field increases with the increment of $a$. Obviously, the   kinetic energy decreases with increasing the scale factor as shown in right panel of Fig. (5). From  this figure   it was observed   that the range of values  of the $\chi(\phi)$ is  $(0.505,0.525)$ from initial  to late time, which is  indicates that the accelerated universe can be obtained for this interval. {\bf Note that the result of Fig. (5)   is in contrast with that obtained by \cite{52b} who showed that the kinetic energy of the  K-essence field  increases  with increasing the scale factor. This difference may come back to this fact that the K-essence field selected by  \cite{52b} is a purely kinetic model in which the action (\ref{71}) is independent of $\phi$.}
\section{COSMOLOGICAL EVOLUTION OF THE INTERACTING HDE and  NADE MODEL IN BI MODEL}
\label{sec5}
In order to fit the model with current observational data, we consider the interacting NADE model in a flat BI universe in this section. In this case, the  Hubble parameter can be written as,
\begin{equation}\label{73}
H=H_0\sqrt{\frac{(1-\Omega_{\Lambda}-\Omega_{\sigma0})X(z)}{1-\Omega_{\sigma0}-\Omega_\Lambda}},
\end{equation}
where $X(z)=\rho_m(z)/\rho_{m0}(z)$.  For the case of FRW ($\Omega_{\sigma0}=0$) which  are in agreement with the respective relations obtained in \cite{53a}. Here $\Omega_{m0}$ is  present value of dimensionless energy density of matter  and  $z$ is the redshift, $z =1/a-1$.  Note that, for simplicity, in this
work we disregard the contributions from baryons and radiation.  For the $\Lambda$CDM model, $H(z)$ as
$H=H_0\sqrt{ \Omega_{m0}(1+z)^3+\Omega_\Lambda }$. Also, for model such as $w$CDM (with
the constant EoS $w$), it is $H=H_0\sqrt{ \Omega_{m0}(1+z)^3+(1-\Omega_{m0}-\Omega_{\sigma0})(1+z)^{3(1+w)}  }$.  The currently preferred values of $w$ in this model is $w=-1.01\pm0.15$ \cite{53}.
 \begin{figure}[tb]
 \includegraphics[width=.6\textwidth]{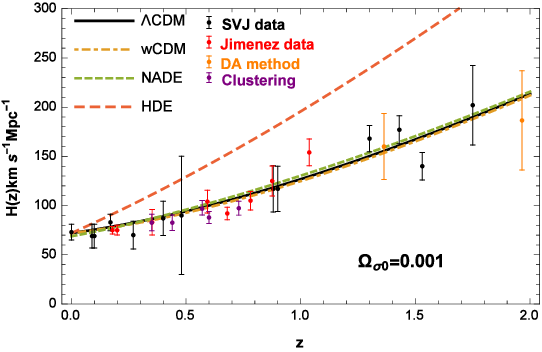}\hspace{.5cm}
 \includegraphics[width=.6\textwidth]{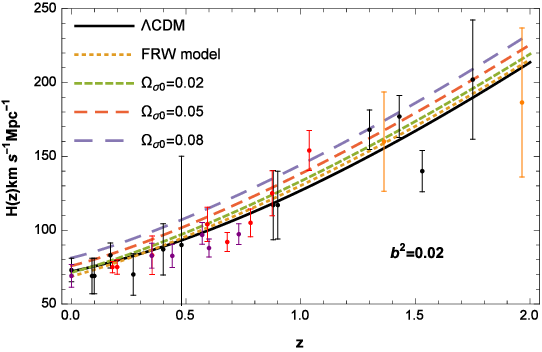}\hspace{.5cm}
 \includegraphics[width=.6\textwidth]{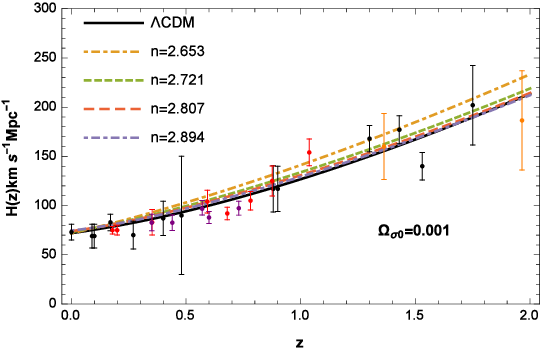}
\caption{Upper panel:  Comparison of the observed  $H(z)$, with the predictions from the NADE, HDE, $\Lambda$CDM and $w$CDM models. Middle panel the evolution of $H(z)$ versus redshift $z$  in NADE anisotropic universe  for different values of    parameter     $\Omega_{\sigma0}$   with  $n=2.886$ and   the $\Lambda$CDM models  while bottom panel the evolution of $H(z)$ versus redshift $z$    for different values of    parameter $n$ with    $\Omega_{\sigma0}=0.001$, by considering  $H_0=72~kms^{-1}Mpc^{-1}$, $\Omega_{m0}=0.274$,  $\Omega^0_\Lambda=0.69$ and $b^2=0.02$. For the case of $\Lambda$CDM model we take $\Omega^0_\Lambda=0.7$ and $\Omega_{m0}=0.3$, and for the $w$CDM model with $\Omega_{m0}=0.3$ and $w=-1.07$ \cite{531}. The data points with errorbars, and theoretical lines for different DE models and the observational $H(z)$ data  \cite{a53,b53,c53,d53}.}
 \label{fig:6}
 \end{figure}
 \begin{figure}[h]
\includegraphics[width=.42\textwidth]{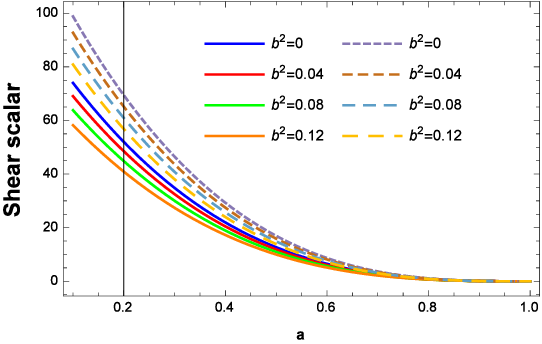}\hspace{1cm}
 \includegraphics[width=.44\textwidth]{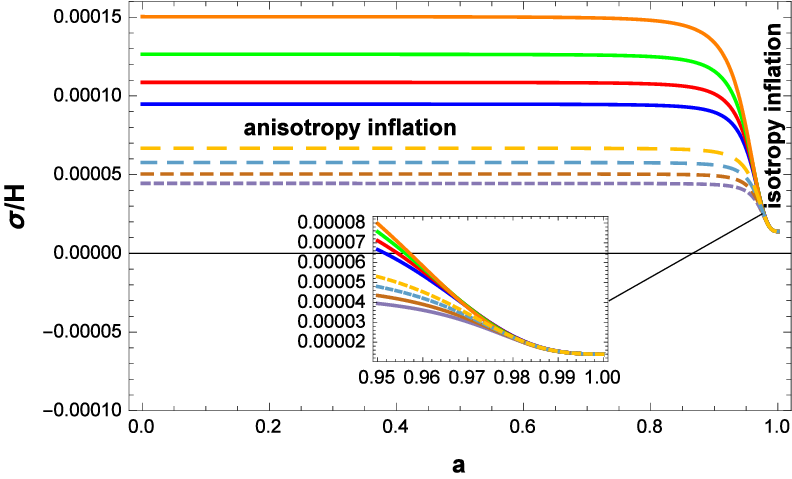}
\caption{Left panel: The evolution of the shear scalar vs. $a$ for   different value of  $b^2$ and $\sigma_0=0.05$ \cite{532}.  Right panel: The evolution of the anisotropy parameter $\sigma/H$ with respect to the  scale factor $a$ in a BI universe.  It becomes almost constant during the anisotropic inflationary phase.  Then it falls rapidly down to $\mathcal{O(\sigma H)}$. As shown in the legend, the  interacting NADE is shown
with the thin lines, and the interacting HDE with the dashed lines.  The data points and error bars are estimated from  Eq.  (\ref{12})  with a fiducial concordance $\Lambda$CDM
model.  The initial data is same as in  Fig. (6).}
 \label{fig:7}
 \end{figure}
 To constrain cosmological parameters we use the $H(z)$ data from SVJ  \cite{a53}, DA method \cite{b53}, Jimenez \cite{c53} and Clustering \cite{d53}. These data, for
the redshift range $0<z<2$, are  shown in figure 6, and we find a remarkable agreement between the different measurements. The value of  $H(z)$ is then directly computed by using Eq.  (\ref{12}).
 In top panel of Fig. (6), we show four models, namely, NADE, HDE,  $\Lambda$CDM and $w$CDM models. We see from this
figure that the cosmic evolution trends are quite similar for   three models, such as, NADE, $\Lambda$CDM and $w$CDM models. So, among these four DE models,
HDE with  the Hubble radius $H^{-1}$  in BI is not   favored by the observational data.
  Also from middle panel of Fig. (6), we can clearly  see that for different  values of $\Omega_{\sigma0}$    the process of cosmic evolution looks quite similar, i.e.,  for the bigger value of the $\Omega_{\sigma0}$  parameter,  the bigger value of the Hubble expansion rate $H(z)$  is  gotten. For this model, we consider, $H_0=72~kms^{-1}Mpc^{-1}$, as the value of  $H_0=73\pm 3 ~kms^{-1}Mpc^{-1}$ from the combination WMAP 3 year estimate \cite{54}, and the other with  $H_0=68\pm 4~ kms^{-1}Mpc^{-1}$ from a median statistics analysis of 461 \cite{55}.
 The evolution of the  $\sigma$ is   plotted in  left panel of Fig. (7). From Fig. (7), it can be observed that evolution of shear depends on the initial conditions. Signature change of shear can be seen for certain kind of  coupling constant $b^2$. Also, the  shear scalar becomes negligible at late times and  $\sigma$ of  NADE decreases slowly for walls compared to the HDE.   Finally we have considered, the model isotropize.
The measure of the anisotropy is described by $\sigma/H$, describe the magnitude of the spacetime shear per   the average expansion
rate.   However, as shown in right panel of Fig. (7), during the inflationary phase, we find that the effect
of shear cannot become as large as the Hubble parameter and  the shear decreases  at the end of the inflation.\\
In the  following, we constrain the model parameters of these   DE models in BI universe by using the $H(z)$ measurements. {\bf In other words, it becomes almost constant during the anisotropic inflationary phase. Then,  at the end of  inflation  $\sigma/H \sim\mathcal{O}(10^{-5})$.}

\subsection{Sandage-Loeb (SL) test}
The redshift-drift observation, sometimes called the {\bf ``Sandage-Loeb (SL) test"}, is not only conceptually simple, but also
is a direct probe of cosmic dynamic expansion, although being observationally challenging.  {\bf The SL test data provide an important supplement to the other DE probes, since they are extremely helpful in breaking the existing parameter degeneracies.}  Ref. \cite{53a} introduced the redshift relation  by   a spectroscopic velocity shift  $\Delta\nu$ as $\Delta\nu\equiv  \Delta z/(1+z)$.  By using the Hubble parameter $H(z)=-\dot{z}/(1+z)$ and Eq.  (\ref{12}), we obtain
\begin{equation}\label{74}
\Delta \nu=H_0\Delta t_0\bigg( 1-\frac{E(z)}{1+z}\bigg),
\end{equation}
where we set $\Delta t_0=10 $ years, and $E(z)=H(z)/H_0$.  According to the Monte Carlo simulations, the uncertainty of $\Delta\nu$ measurements expected by CODEX can be expressed as \cite{55a}
\begin{equation}\label{a74}
\sigma'_{\Delta \nu}=1.35(\frac{2370}{S/N})\sqrt{\frac{30}{N_{QSO}}}(\frac{5}{1+z_{QSO}})^{1.8} cm/s,
\end{equation}
where $S/N=3000$ is the signal-to-noise ratio, $N_{QSO}=30$ is the number of observed quasars and $ z_{QSO}\in [2,5]$ is their redshift. {\bf  The fiducial concordance cosmological model with the parameters taken to be the best-fit ones from WMAP nine years analysis \cite{55aa} is applied to examine the capacity of future measurements of SL test to constrain the concerned models. A consistency test of the geometric and structural measurements
might provide a diagnostic to the origin of the acceleration of the universe in the future.
Of course, the SL test will definitely play a significant role in doing such an analysis.}   We will examine the SL test, and then examine effects of anisotropy and parameter of $n$ on the NADE in   the  SL test as shown in Fig. (8). We have chosen the fractional matter density $\Omega_{m0}=0.274$ from $\Lambda$CDM  \cite{30} and $n = 2.807$ \cite{52}. {\bf We  also use the $\Lambda$CDM model as a fiducial model to perform an SL test.}  In  left panel of Fig. (8), the parameter $n$ is fixed to be $2.807$  for the NADE model, and the parameter $\Omega_{\sigma0}$ of the NADE model is variable while in right panel, we fix $\Omega_{\sigma0}=0.001$ and vary $n$ for the NADE model in BI universe. Based on the parameter spaces constrained from the current data
combination, the boundaries of  $\Delta\nu$ could be determined by using Eq. (\ref{74}). We also plot
the error bars in the SL test, given by Eq. (\ref{a74}), on the corresponding bands, in order to
make a direct comparison with the reconstructed results from the current data.
From Fig. (8) we see that the NADE  in BI model can be distinguished from the $\Lambda$CDM and $w$CDM  models via the  SL  test. As we can observe
$\Delta \nu$ is positive at small redshifts and becomes negative at $z\geq1.7$. This result is reasonable with Refs. \cite{53a,531,56}. Clearly, for different values of $\Omega_{\sigma0}$ and $n$, low\textbf{-}redshift probes cannot distinguish these different models since the required redshift is too high even for the
most ambitious surveys.  Beside,  the amplitude and slope of the signal depend mainly on both  $\Omega_{\sigma0}$ and $n$. {\bf  So, we see that the prospective SL test is very powerful to be used to constrain the NADE  of BI model, and it is clearly better than the current low-redshift observations.}
\begin{figure}
\includegraphics[width=0.46\textwidth]{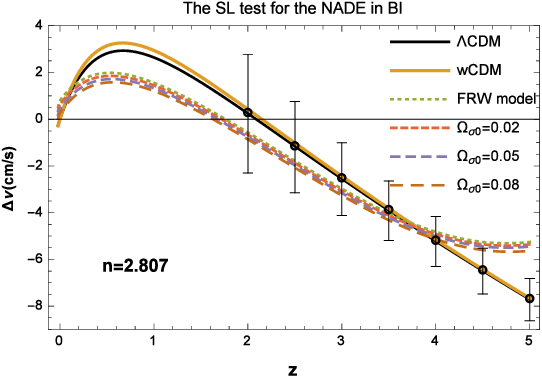}\hspace{.1cm}
 \includegraphics[width=.46\textwidth]{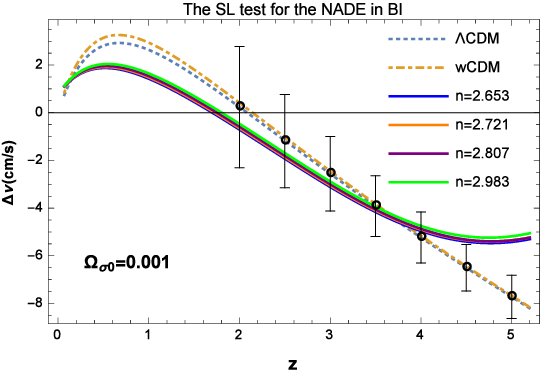}
\caption{Cosmic velocity shift as function of the   redshift for the NADE model with four different values of      $\Omega_{\sigma0}$ (left panel) and $n$ (right panel),  the $\Lambda$CDM  model with  $\Omega^0_{\Lambda}=0.7$  and $w$CDM model with $w=-1.07$ and  $\Omega_{m0}=0.3$. A time
interval of 10 years has been assumed. The   data points and error bars are estimated from Eq. (\ref{a74}) with a fiducial concordance $\Lambda$CDM model.}
\label{fig:8}
\end{figure}

\subsection{The statefinder diagnostic}
\begin{figure}[tb]
 \includegraphics[width=.35\textwidth]{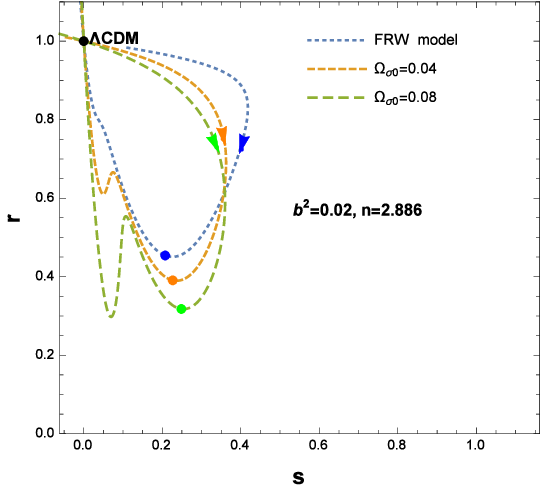}\hspace{1.5cm}
 \includegraphics[width=.35\textwidth]{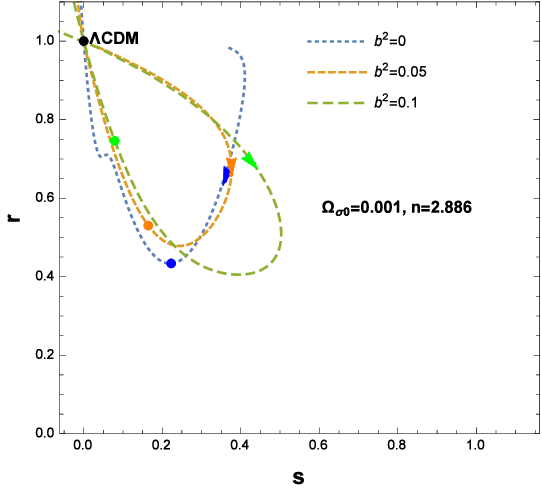}
  \includegraphics[width=.35\textwidth]{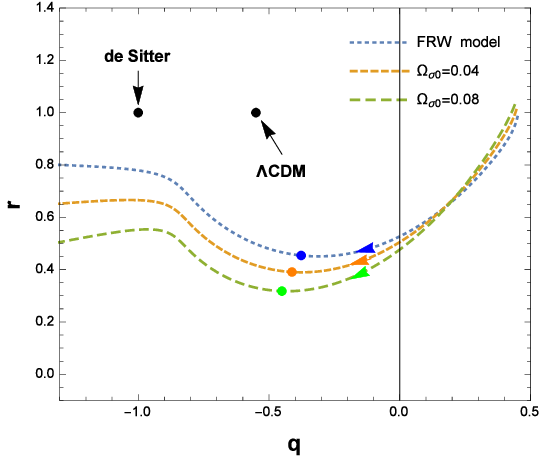}\hspace{1.5cm}
 \includegraphics[width=.35\textwidth]{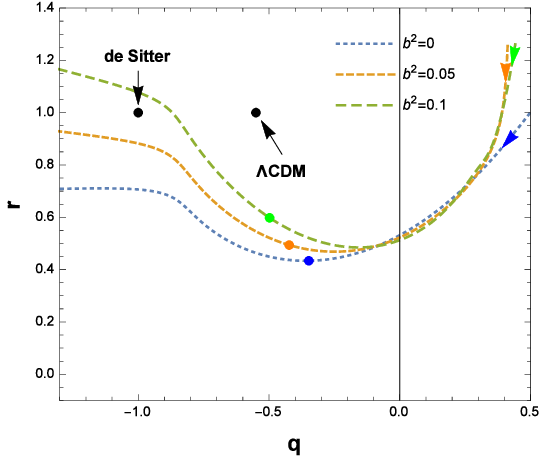}
  \caption{ Top panel: The statefinder diagrams $r(s)$ for the interacting NADE  in BI model. The $\Lambda$CDM model corresponds to a fixed point $\{0, 1\}$. The dots show today's values for the statefinder parameters $(s_0, r_0)$. Bottom panel: Evolutionary trajectory in the $r-q$ plane for interacting NADE model in BI universe.}
 \label{fig:9}
 \end{figure}
In order to achieve a strong analysis to discriminate among  DE models,  Sahni \textit{et al}. \cite{56a} have
presented a new geometrical diagnostic pair  $\{r, s\}$, known as statefinder parameters, which is constructed from the scale factor and its derivatives up to the third order. The statefinder pair is a ``geometrical" diagnostic in the means that it is constructed from a space-time metric directly. In the following, we consider the statefinder parameters $\{r, s\}$ \cite{56a,56b,56c,56d} for the present case. The ${r, s}$ parameters are given by \cite{56a}
\begin{eqnarray}\label{74a}
r\equiv   \frac{\dddot{a}}{aH^3},\quad~~s\equiv\frac{r-1}{3(q-\frac{1}{2})}.
\end{eqnarray}
 We will soon see that it has a remarkable property for the basic flat
$\Lambda$CDM BI model. The statefinder parameter $s$ is a linear combination of $r$ and $q$.  In the $\{r, s\}$ plane, $s>0$ corresponds to a quintessence-like model of DE and $s<0$ corresponds to a phantom-like model of DE.  In addition, other
proposals on the statefinder diagnostic  include varies DE models  \cite{56e,56g,56h}.
Then, by using the BI equation, we can obtain the following concrete expressions
\begin{eqnarray}\label{74b}
&&r=1+\frac{3}{2}\Omega_{\sigma 0}+\frac{9}{2}\frac{\Omega_{\Lambda}\omega_{\Lambda}}{1-\Omega_{\sigma 0}}(1+\omega_{\Lambda}+\Omega_{\sigma 0})+\frac{9}{2}\omega_{\Lambda}b^2 -\frac{3}{2}\frac{\Omega_{\Lambda}\omega'_{\Lambda}}{1-\Omega_{\sigma 0}},\cr
&&s=\frac{\Omega_{\sigma 0}+\frac{3\Omega_{\Lambda}\omega_{\Lambda}}{1-\Omega_{\sigma 0}}(1+\omega_{\Lambda}+\Omega_{\sigma 0})+3\omega_{\Lambda}b^2 -\frac{\Omega_{\Lambda}\omega'_{\Lambda}}{1-\Omega_{\sigma 0}}}{3\omega_{\Lambda}\Omega_{\Lambda}-\Omega_{\sigma0}},
\end{eqnarray}
where $\Omega_{\Lambda}$ is given by  (\ref{61}) and $\omega'_{\Lambda}=\bigg(\frac{\sqrt{\Omega_{\Lambda}}}{na}+\frac{3b^2(1-\Omega_{\sigma0})}{\Omega_{\Lambda}}\bigg)\bigg(1-b^2-\frac{\Omega_{\Lambda}}{1-\Omega_{\sigma0}}-\frac{2}{3na}(1-\frac{\Omega_{\Lambda}}{1-\Omega_{\sigma0}})\sqrt{\Omega_{\Lambda}}\bigg)$. So by solving Eq. (\ref{61}) we can get the evolution solution of  $\Omega_{\Lambda}$
and then hold all the cosmological quantities of interest and the whole dynamics of the universe.  The $\{r, s\}$ evolutionary trajectory in the interacting NADE model for different values of $\Omega_{\sigma0}$ and $b^2$ are shown in top panel of Fig. 9. The arrows
in the diagram denote the evolution directions of the statefinder trajectories and the color dots show present-day values for the statefinder parameters. The crossing with  $\Lambda$CDM  occurs for this case.
 It can be seen that SCDM   and  $\Lambda$CDM models  have fixed point value of statefinder
pair  $\{r, s\}=\{1, 1\}$ and  $\{r, s\}=\{1, 0\}$, respectively. From Fig. 9, we can find that the  the larger anisotropy parameter   makes the value of $r$ smaller and the value of $s$ bigger while the  the larger interaction between dark components makes the value of $s$ smaller and the value of $r$ bigger, evidently. From this figure one can observe that, in the low-redshift region, the difference
between the NADE models and the $\Lambda$CDM model can be easily distinguished, which is quite different from the cases of $H(z)$ (see Fig. 6).
{\bf The $r-s$ planes shows the NADE behavior,   quintessence  limit  and will also approaches to $\Lambda$CDM behavior (see top panels of Fig. 9)}.  As a complementarity, bottom panel of Fig. 9 shows another statefinder diagram $r-q$ evolutionary trajectory.  {\bf In this case, we clearly observe that both $\Lambda$CDM ( $r=1,q=-0.55$) scenario and NADE model commence evolving from the same point in the past $r=1,q=0.5$ which corresponds to a matter dominated SCDM universe, and end their evolution at the same common point in the future  $r=1,q=-1$ which corresponds to  the de Sitter expansion}.  In Table 1  we have computed different values of $r, s, q$ for the current universe $(a = 1)$ and for different choices of $b^2$ and $\Omega_{\sigma0}$.\\
\hspace{3mm}{\small {\bf Table 1.}} {\small ~~~~
Today's values of $r$, $s$, $q$ with $n=2.886$.}\\
    \begin{tabular}{l l l l l l l  p{0.15mm} }
    \hline\hline
  \vspace{0.50mm}
{\footnotesize  Statefinder parameters } &  {\footnotesize    $~~~b^2=0 $ } & {\footnotesize  $~~~ b^2=0.05$ } & {\footnotesize $~~~b^2=0.1$ } &
 {\footnotesize  $~~~\Omega_{\sigma0}=0$ } & {\footnotesize  $~~~\Omega_{\sigma0}=0.04$ }  & {\footnotesize  $~~~\Omega_{\sigma0}=0.08$ }  \\\hline

\vspace{0.5mm}

{\footnotesize $~~~~~~~~~~~~~ r$} &
{\footnotesize  $~~~~0.433 $ } & {\footnotesize  $~~~~0.530$ } & {\footnotesize  $~~~~0.745$ } &
{\footnotesize  $~~~~0.454$ }  & {\footnotesize  $~~~~0.391$ } & {\footnotesize  $~~~~0.317$ } \\

\vspace{0.5mm}

{\footnotesize $ ~~~~~~~~~~~~ s$} &
{\footnotesize  $ ~~~~0.222 $ } & {\footnotesize  $~~~~0.164$ } & {\footnotesize  $~~~~0.079$ } &
{\footnotesize  $~~~~0.207$ }  & {\footnotesize  $~~~~0.226$ } & {\footnotesize  $~~~~0.248$ } \\

\vspace{0.5mm}

{\footnotesize $~~~~~~~~~~~~q $} &
{\footnotesize  $ ~~-0.348 $ } & {\footnotesize  $~~-0.423$ } & {\footnotesize  $~~-0.498$ } &
{\footnotesize  $~~-0.377$ }  & {\footnotesize  $~~-0.412$ }  & {\footnotesize  $~~-0.451$ }\\

\hline

\vspace{0.5mm}
 \end{tabular}
 \vspace{5mm}

\subsection{Linear perturbation theory}
\begin{figure}[h]
\centerline{\includegraphics[width=.4\textwidth]{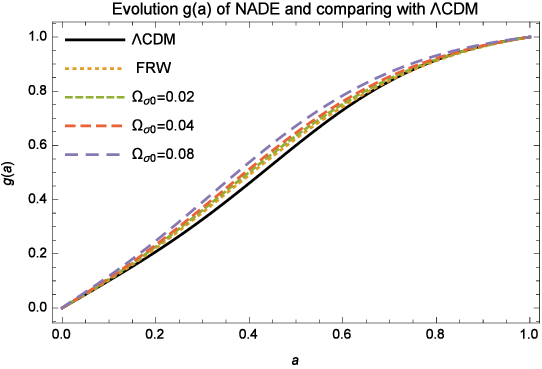}}
\caption{Time evolution of the growth factor for  different value of the anisotropy energy density parameter $\Omega_{\sigma0}$ and comparing to the $\Lambda$CDM and FRW in NADE models with $n=2.807$.  }
 \label{fig:10}
 \end{figure}
In this subsection, we study the linear growth of perturbation of non relativistic dust matter by computing the evolution of growth factor $g(a)$  in NADE  model, and then compare it with the evolution of growth factor in  NADE of FRW and $\Lambda$CDM models.  In this case the differential equation for the evolution of $g(a)$ is
given by \cite{57,58}
 \begin{equation}\label{75}
g''(a)+\bigg(\frac{3}{a}+\frac{E'(a)}{E(a)}\bigg)g'(a)-\frac{3}{2}\frac{\Omega_{m0}}{a^5E^2(a)}g(a)=0,
\end{equation}
{\bf where the prime denotes a derivative w.r.t. the scale factor.} In order to study the linear growth in NADE model, using Eqs.  (\ref{56}),  (\ref{61}) and  (\ref{73})  for  BI universe,
we solve numerically Eq.  (\ref{75}).  In addition, we solve numerically  (\ref{75}) for the FRW model and the $\Lambda$CDM model. To evaluate the initial conditions, since we are in the linear regime, we assume that the linear growth factor has a power law solution, $g(a)\propto a^m$ \cite{57}, with $m$, to be evaluated at the initial time.
We plot the evolution of $g(a)$  with respect to a function of the scale factor in Fig. (10). In the $\Lambda$CDM model, the growth factor evolves more slowly compared to
the NADE in FRW universe because the expansion of the universe slows down the structure formation, \textbf{but at the late time the curves to be coincide because negligible the effect of anisotropic  energy density parameter  }. Also, in the NADE of FRW model, the growth factor evolves more slowly compared to the NADE in BI model. This behavior can be explained by taking into account the evolution of Hubble parameter in Fig. (6).
Therefore the growth factor $g(a)$ for the  $\Lambda$CDM and DE of FRW model will always fall behind the NADE in anisotropic universe.

\subsection{ Distance modulus }
 Finally, we consider constraints on model parameters coming from SNIa observations. Observation of SNIa does not provide standard ruler but rather gives distance modulus. This quantity is defined by  \cite{59}
\begin{equation}\label{76}
\mu_{th}(z)=5 \log_{10}\frac{D_L(z)}{Mpc}+25.
\end{equation}
The luminosity distance $D_L(z)$ is given by
\begin{equation}\label{77}
D_L(z)=(1+z) r(z),
\end{equation}
where $r(z)=\int ^z_0{ H^{-1}(z')}dz'$.   In all, current SNIa data are unable to discriminate between the popular $\Lambda$CDM and our interaction model. In Fig. 11, we show the  dimensional marginalized contour in the  $c-\Omega_{\sigma0}$   for the HDE model in an anisotropic universe. The fit values for the model
parameters is $c=0.994^{+0.126}_{-0.065}$ at the moment time i.e. $z\rightarrow 0$. As is clear from figure 11 the  parameter $c$ increases with the decreasing anisotropy parameter
and this is consistent with the observed data.  We can also measure $D_L(z )$ through the Hubble parameter by using the Eqs.  (\ref{32}) and  (\ref{73}).
Figure 12 presents the distance modulus with the best fit of our model and the best fit of the $\Lambda$CDM model to the Gold and Silver SNe Ia  data sets of \cite{60,61}.  We add 4 SNe Ia from this current paper. From Fig. 12 we can observe the universe is accelerating expansion. In all, current data are unable to discriminate between the popular $\Lambda$CDM, FRW and our interaction models as shown in   Fig.   (12).
 \begin{figure}[tb]
\centerline{ \includegraphics[width=.3\textwidth]{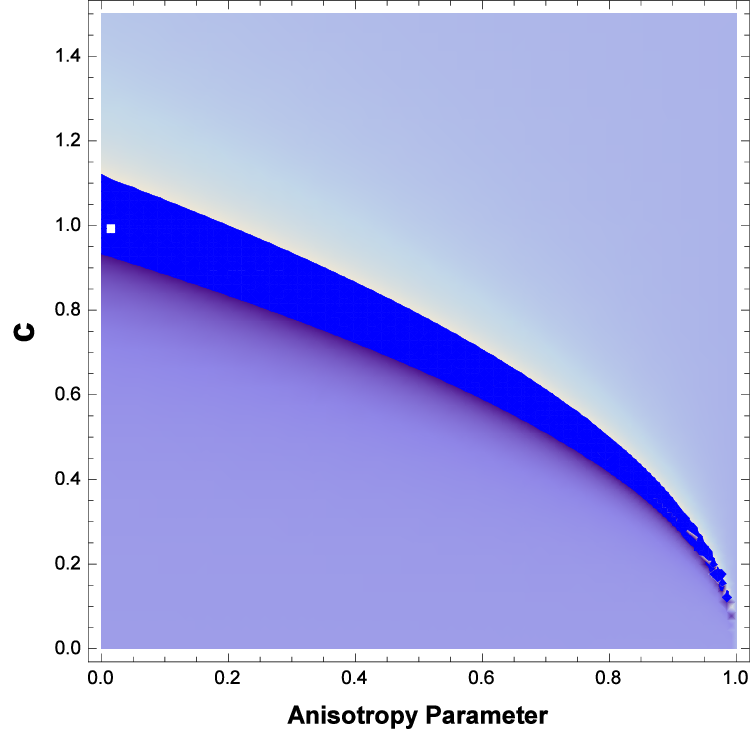}}
\caption {The constraints on the HDE model in the $c-\Omega_{\sigma0}$ plane from
the current $H(z)$ measurements in an anisotropic universe.}
 \label{fig:11}
 \end{figure}
\begin{figure}[tb]
 \centerline{\includegraphics[width=.4\textwidth]{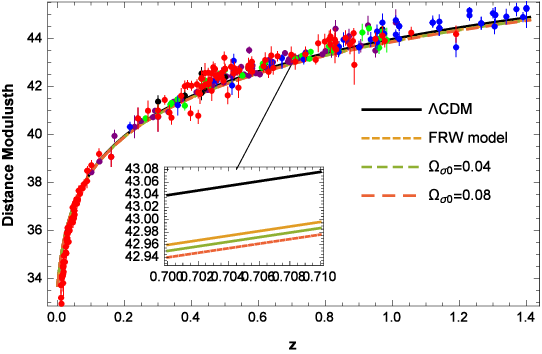}}
  \caption{ Distance modulus   for the best fit model $\Omega_{m0}=0.288$, $H_0=72~km/s/Mpc$,  $b^2=0.02$, $n=2.807$ and   the $\Lambda$CDM model,  $\Omega_{m0}=0.3$, $H_0=72~km/s/Mpc$ and $\Omega^0_{\Lambda}=0.7$. The data points are
from the gold and Silver SNe Ia  of \cite{60,61}.}
 \label{fig:12}
 \end{figure}

\section{Conclusion and Discussion}
\label{sec6}

In this work, we have investigated the different models of cosmology in an anisotropic universe,\textbf{with resemblance between their energy density to scaler fields energy density.}
 In particular, \textbf{we have reconstructed   the dynamics i.e. evolution equations of these scalar field models according to the evolutionary behavior  dark energy models in an anisotropic universe.}
 Then we have reconstructed  both the potentials and the kinetic energies corresponding to each model, which describe quintessence, tachyon and  K-essence
cosmology.   We have discussed the effects of the interaction and anisotropy on the evolutionary behavior
the  HDE and NADE  scalar field models. {\bf Our main conclusions can be summarized as follows.}
\begin{itemize}
\item
The simplest choice is the Hubble scale $L=1/H$, giving a energy density that is comparable to the present-day DE  \cite{9,451}.  It has been argued that an IR cutoff
defined by the Hubble radius, \textbf{cause} the DE regard as pressureless so can not lead to an accelerated universe. We have also shown that taking into account the  interaction  \cite{42,45} and anisotropy terms, we are able to describe the accelerating universe.  In the following, by using  the Hubble radius as    $L=H^{-1}$ IR cutoff,  combinations of the HDE model and interacting scalar fields in anisotropic universe were implemented.  We have obtained expressions for the scalar field and the  potential for interacting HDE in BI model.  The free parameters of cosmology have been  evaluated  in terms of the constants $b^2$, $\Omega_{\sigma0}$ and $c^2$. The results of
our analysis are displayed in Fig. (11).  We used current observational data to constrain
the HDE model. The fit values for the model
parameters is $c=0.994^{+0.126}_{-0.065}$ at the moment time.
\item
The EoS parameter $\omega_\Lambda$ of the NADE model in  the BI models,   can   cross the phantom divide line ($\omega_\Lambda<-1$) at the present time, provided  that  $b^2\geq0.08$ which is compatible with the observations. By the way, the effect of various $\Omega_{\sigma0}$ were negligible but as $a\rightarrow 0$, it was clear that decreasing of the $\Omega_{\sigma0}$ cause to decrease the EoS parameter as shown in left panel of figure 1.
{\bf The evolution of   the interacting NADE density parameter  $\Omega_{\Lambda}$ is depend on the anisotropy density parameter $\Omega_{\sigma0}$.in Figure (2a)  it was clear that $\Omega_{\Lambda}$ increase with increase of anisotropic parameter density at the smaller scale factor, but at the $a=1$, the effect of anisotropic parameter density is negligible,that is in agreement with observational data, i.e. the present universe is close to  homogeneous and isotropic flat universe.  Figures (2b) and (2c)  evolution of $\Omega_{\Lambda}$ have plotted for different choices of $b^2$ and $n$,it is clear,  the $b^2$  parameters increase and the $n$ decrease  cause the $\Omega_{\Lambda}$ increase at the smaller scale factor but at the $a=1$, the curves are coincide, that is also in agreement with the observational data.}
\item
The new agegraphic quintessence  scalar {\bf field} for a given $\Omega_{\sigma0}$ increases with increasing the scale factor; also, the curve was shifted to the larger value of $\phi$ with decreasing $\Omega_{\sigma0}$ as shown in left panel of Fig.  (3). For a given $\Omega_{\sigma0}$, potential $V(\phi)$ decrease with
increasing the scale factor and the curve is shifted to the bigger values of $V(\phi)$ with increasing the $\Omega_{\sigma0}$ as shown in figure 3 (middle panel).
  Right panel of Fig. (3) shown which the fraction of potential per kinetic energy increase with the $a$ increase and its magnitude increase with the $\Omega_{\sigma0}$ decrease for the smaller $a$. In addition, that is consist of with slow-roll approximation on  inflation era for the smaller a at early
universe.

\item
The new agegraphic  tachyon scalar {\bf field}   for a given  $\Omega_{\sigma0}$ behave like the new agegraphic  quintessence model as shown in figure 4.
\item
We have evaluated   the interacting NADE versions of K-essence scalar field in BI models.  These results have been shown in Fig. (5). The  new agegraphic K-essence  scalar field for a given  $\Omega_{\sigma0}$ increases with increasing the scale factor. But its kinetic energy decreases. For a given scale factor, the
new agegraphic K-essence   scalar {\bf field}  and kinetic energy decreases with increasing  $\Omega_{\sigma0}$. Anyway,  we have shown that the evolution of the NADE   scalar {\bf field} and  kinetic energy in anisotropic universe is smaller than the  evolution of the NADE   scalar {\bf field} and  kinetic energy in   FRW space time. But its potential   large.
\item
Considering a  BI model and interacting between DE and DM we have reconstructed Hubble parameter $H(z)$ (see  Fig. (6)) versus   the redshift  $z$ and  it \textbf{was} compared the observed expansion rate $H(z)$ with that predicted by the  four DE models. We have used the SVJ, DA method   and Jimenez  Hubble parameter versus redshift data to constrain cosmological parameters of different DE models. The constraints are restrictive, and consistent with those determined by using type Ia supernova redshift-magnitude data.   The reconstructed Hubble parameter is found to be decreasing with evolution of the universe   and according to this model, the evolution of the accelerated expansion of the universe is speedy than the evolution of the isotropic model. It has seen that among different  DE models, only the curve of $H(z)$ has predicted by the NADE model.  So,  HDE with the Hubble radius  $1/H$ in BI is not favored by the observational data. We have analyzed the evolution of the shear scalar in right panel of  Fig. (7), which was depending on the initial conditions  and we found that there can be a \textbf{ change of sign} in the shear. This may have some interesting implications in early universe cosmology.
Furthermore, we have studied  the evolution of the  $\sigma/H$ as shown in  left panel of figure 7.
Although the shear is constant during the inflation, it can  fall   at the end of the inflation. Moreover, we have seen that \textbf{fraction of} the anisotropy of the universe \textbf{to shear scaler}
$\sigma/H$ can decrease to $\mathcal{O}(10^{-5})$ at the end of the inflation. {\bf  The Sandage-Loeb (SL) test   directly measures the temporal variation of the redshift of distant quasars (QSO)  Lyman-$\alpha$ absorption lines in the so-called ``redshift desert" $(2\lesssim z \lesssim5)$, which is not
covered by any other cosmological observation.  In Refs. \cite{61a,61b,61c}, they were performed a serious synthetic
exploration of the impact of future SL test data on DE constraints. It was shown that the SL test can break the parameter degeneracies in existing DE probes and significantly improve the precision of DE constraints.}  In the following,  we   analyzed how the   SL test  would impact on the DE constraints from
the future geometric measurements. Therefore,  the   SL test  can be used to distinguish the NADE of FRW and  NADE of BI models from the $\Lambda$CDM, the $w$CDM and  it was observed that the constraints on  $\Omega_{\sigma0}$  and $n$ are very strong (see Figs. (8)).
   Comparing the dynamical {\bf behavior} of the expansion expected in anisotropic universe  with isotropic universe.
  {\bf  The statefinder parameters  are expected to be useful tools in
testing interacting cosmologies that solve or at least alleviate the coincidence problem which
besets many approaches to late acceleration. For this purpose, we applied} the statefinder diagnostic to the NADE model   with interaction in BI model.  The statefinder diagrams show that the interaction between DE and DM can significantly affect the evolution of the universe and the contributions of the interaction can be diagnosed out explicitly in this method. We plot the evolutionary trajectories of this model in the statefinder
parameter planes as shown in Fig. (9). The statefinder diagnostic can   be performed to the NADE model in cases of different  $\Omega_{\sigma0}$ and $b^2$, which indicates that the value  of $\Omega_{\sigma0}$ and $b^2$ determines the evolution behavior and the fate of the universe.  The BI NADE model can  reach the $\Lambda$CDM point $\{1, 0\}$ in the $r-s$ plane. In the limiting case of $b^2=0.02$, the trajectory of $r-s$ plane with all values of $\Omega_{\sigma0}$ approaches to quintessence  behavior as well as corresponds to $\Lambda$CDM limit as shown in top panel of Fig. (9). It is also valid for $\Omega_{\sigma0}=0.001$ with different values  of $b^2$.  In addition, the $r-q$ plane has been used for discussion on the evolutionary property of the BI universe (see bottom panel of Fig. (9)).
{\bf We have started our analysis by studying the effects of the anisotropy of the background expansion history of the growth factor and
on a second step we have  taken into account also the perturbations in the NADE component.   We showed that the linear growth factor
is sensitive to the details of the model considered. In particular, } it is observed that   the growth factor $g(a)$ for the $\Lambda$CDM and DE of FRW model will always fall behind the NADE in an anisotropic universe (see  figure (10)).  In addition to this we discuss  the distance modulus with the best fit of our model and the best fit of the $\Lambda$CDM model to the Gold and Silver SNe Ia  data sets of \cite{60,61}. In  Fig. (12) we  are able to discriminate between the  $\Lambda$CDM, FRW and NADE in BI model.  {\bf As a result, the NADE of BI model is consistent with current observations and the more precise cosmological observations will be taken to be the decided
constraints on this model.} \\
Finally, we would like to mention that the aforementioned discussion in first part  this work can be easily generalized to other choices of IR cut-off, namely, Ricci length scale and radius of the event horizon (for this point see also, e.g.  \cite{62}).
\end{itemize}
{\bf  Acknowledgements }  The authors are grateful to the referee for valuable comments and suggestions that have allowed us to improve this paper significantly.


\end{document}